\documentclass{article}

\PassOptionsToPackage{numbers,sort&compress}{natbib}
\usepackage[preprint]{neurips_2026}

\usepackage[utf8]{inputenc} 
\usepackage[T1]{fontenc}    
\usepackage{hyperref}       
\usepackage{url}            
\usepackage{booktabs}       
\usepackage{amsfonts, amsmath,amssymb}       
\usepackage{nicefrac}       
\usepackage{microtype}      
\usepackage{xcolor}         
\usepackage{listings}
\usepackage{enumitem}
\usepackage{tikz}
\usepackage{geometry}

\usepackage{adjustbox}
\usetikzlibrary{arrows.meta,positioning,shapes.geometric,fit,backgrounds}
\usepackage{microtype}

\hypersetup{colorlinks=true,linkcolor=blue,citecolor=blue,urlcolor=blue}
\usepackage{algorithm}
\usepackage{algpseudocode}

\usepackage[dvipsnames]{xcolor} 
\usepackage{tcolorbox} 

\newtcbox{\badge}[1][gray]{
  on line, 
  arc=4.5pt, 
  colback=#1!80!black,
  colframe=#1!80!black,
  before upper=\strut, 
  boxrule=0pt, 
  boxsep=0pt, 
  left=2.5pt,
  right=2.5pt, 
  top=1.5pt,
  bottom=0.3pt,
  baseline=2pt, 
  fontupper=\fontsize{7pt}{8pt}\selectfont\color{white}\bfseries
}
\usepackage[svgnames]{xcolor}
\usepackage{caption}
\definecolor{CustomPurple}{HTML}{9683FF}
\definecolor{CustomGreen}{HTML}{01810E}
\definecolor{CustomOrange}{HTML}{E76F08}

\newcommand{\mechanismA}{\badge[CustomPurple]{\textbf{A}}}
\newcommand{\mechanismB}{\badge[CustomGreen]{\textbf{B}}}
\newcommand{\mechanismC}{\badge[CustomOrange]{\textbf{C}}}
\title{
  Continuous Discovery of Vulnerabilities in LLM Serving Systems with Fuzzing 
}

\author{%
  Yunze Zhao \\
  University of Maryland \\
  \texttt{yunzez@umd.edu} \\
  \And
  Yibo Zhao \\
  University of Maryland \\
  \texttt{yibozhao@umd.edu} \\
  \AND
  Yuchen Zhang \\
  New York University \\
  \texttt{yzhang0701@gmail.com} \\
  \And
  Zaoxing Liu \\
  University of Maryland \\
  \texttt{zaoxing@umd.edu} \\
  \And
  Michelle L. Mazurek \\
  University of Maryland \\
  \texttt{mmazurek@umd.edu} \\
}

\begin{document}
\lstset{
  basicstyle=\ttfamily\footnotesize,
  breaklines=true,
  frame=single,
  backgroundcolor=\color{gray!10},
  keywordstyle=\color{blue},
  commentstyle=\color{green!50!black},
}

\maketitle

\begin{abstract}

LLM inference and serving systems have become security-critical infrastructure; however, many of their most concerning failures arise from the serving layer rather than from model behavior alone. Modern inference engines combine KV cache, batching, prefix sharing, speculative decoding, adapters, and multi-tenant scheduling, creating shared-state behavior that only emerges under realistic concurrent workloads and is missed by standard model, safety, and API tests. We present GRIEF, a greybox fuzzer for LLM inference engines that treats timed multi-request traces as first-class inputs, uses lightweight oracles to detect crashes, hangs, performance pathologies, and silent output corruption, and applies controlled replay with log-probability checks to confirm reproducible serving-layer failures. Across early campaigns on vLLM and SGLang, GRIEF discovers 15 vulnerabilities, 10 confirmed by engine developers, including 2 CVEs, spanning KV-cache isolation failures, cross-request performance interference, and crash or liveness bugs. These results show that concurrency, caching, and state reuse can induce silent cross-request contamination, noisy-neighbor denial of service, and delayed crashes without malformed inputs or explicit server errors, making concurrent serving behavior a first-class security and reliability boundary for LLM infrastructure.

\end{abstract}

\newcommand{\mynote}[3]{{\color{#1}\bfseries [#2: #3]}}

\ifdefined\disablecomments
\newcommand{\yunze}[1]{}
\newcommand{\michelle}[1]{}
\newcommand{\yibo}[1]{}
\newcommand{\alan}[1]{}

\else

\definecolor{beige}{rgb}{0.82, 0.71, 0.55}
\newcommand{\yunze}[1]{\mynote{blue}{Yunze}{#1}}
\newcommand{\yibo}[1]{\mynote{purple}{Yibo}{#1}}
\newcommand{\alan}[1]{\mynote{red}{Alan}{#1}}
\newcommand{\michelle}[1]{\mynote{beige}{Michelle}{#1}}
\newcommand\yuchen[1]{\textcolor{orange}{Yuchen: #1}}

\newcommand\draft[1]{\textcolor{cyan}{Draft: #1}}

\newcommand{\code}[1]{{\fontfamily{lmtt}\selectfont{#1}}}

\section{Introduction}
LLM-inference and model-serving systems, such as vLLM~\cite{VLLM2025}, SGLang~\cite{sglang}, and llama.cpp~\cite{GgmlorgLlamacpp2026}, have become core infrastructure for modern AI applications. These systems implement aggressive optimizations, including KV caches, dynamic batching, prefix sharing, and speculative decoding~\cite{VLLM2025,
orca2022,sglang,specDecode2023}, to meet the compute and memory demands of large models. 
As a result, the serving stack itself has become a security-critical boundary: it determines which requests share cached state, which tenants co-occupy a batch, which LoRA adapters are active, and how execution state is reused over time. This complexity makes inference engines highly bug-prone and hard to test, and faults in the serving systems can manifest as crashes, hangs, corrupted shared state, output perturbations, or latent performance anomalies rather than clean failures. In fact, recent empirical analysis~\cite{liu2025first} indicates that more than 35\% of bugs manifest as non-crash anomalies 
rather than outright crashes.

This paper presents the first early evidence that LLM serving systems expose a distinct and under-tested vulnerability surface. Moving away from model security, such as prompt-level jailbreaks~\cite{chao2024jailbreakbench, zou2023universaltransferableadversarialattacks, liu2024autodangeneratingstealthyjailbreak, chao2024jailbreakingblackboxlarge, mazeika2024harmbenchstandardizedevaluationframework}, model hallucinations~\cite{lam2025codecrash, kuhn2023semantic, tian2024finetuning}, or ordinary API-compliance bugs~\cite{luo2021graph}, we study failures arising from the software execution logic of otherwise valid inference requests. Specifically, we demonstrate several significant failures, including a concurrent workload causing one request to reuse stale KV-cache state from another, allowing cross-request output contamination; a valid but adversarially constructed request externalizing its cost onto unrelated co-scheduled users, causing severe first-token latency and near-zero useful throughput without crashing the server; and a batch of individually-valid adapter-serving requests violating a scheduler invariant and terminating the serving process. 
These failures are concerning because they can occur without malformed inputs, explicit server errors, or obvious model-quality degradation.


We believe these behaviors should be treated as an early warning for LLM infrastructure security. 
Existing testing approaches~\cite{chao2024jailbreakbench, deng2023large, luo2021graph, fioraldi2020aflpp,yu2026enabling,  LibAFLFuzzingLibrary} do not naturally express this threat mode. Model-level evaluations test whether a model provides answers safely or correctly~\cite{chao2024jailbreakbench}. API tests~\cite{luo2021graph} check whether individual requests are handled correctly. Conventional fuzzers~\cite{fioraldi2020aflpp, fioraldi2022libafl} are effective at finding crashes from malformed or low-level inputs. None of these approaches systematically exercise the timing, overlap, and shared-state interactions that determine whether a serving system preserves isolation under realistic concurrent workloads.

To systematically test this attack surface, we build \textbf{GRIEF} (\textbf{GR}eybox \textbf{I}nference \textbf{E}ngine \textbf{F}uzzer), a greybox fuzzer for LLM inference and serving systems. Here, ``greybox'' indicates that GRIEF does not treat the server as a pure opaque box: it observes lightweight execution feedback, such as latency, request outcomes, resource usage, and KV-cache events, and uses that feedback to guide which concurrent request traces to mutate and replay. GRIEF treats a concurrent client workload as the fuzzing input. Its core abstraction is a timed \emph{request trace}: a sequence of events that captures not only what clients ask, but when requests overlap and how they compete for shared serving state. GRIEF mutates these traces to explore co-batching, prefix-cache reuse, adapter co-scheduling, cancellation timing, and scheduler pressure. Traces let GRIEF search for failures that only emerge from workload structure rather than from any single request in isolation.

Detecting these failures requires oracles beyond simple crash detection. GRIEF therefore combines behavioral, structural, and relational oracles. Behavioral oracles detect externally visible anomalies such as severe latency amplification or corrupted outputs. Structural oracles use platform-specific telemetry, such as KV-cache events, to identify suspicious state reuse or resource-management anomalies. Relational oracles compare executions across related \emph{prompt families}, allowing GRIEF to distinguish benign decoding variation from serving-layer divergence. To address non-determinism in LLMs, GRIEF uses a two-stage confirmation architecture: suspicious executions are deferred to controlled replay, where majority-vote confirmation and token log-probability (logprob) checks separate reproducible bugs from infrastructure noise.

Our early campaigns demonstrate that this threat model is not hypothetical: by repeatedly mutating and executing concurrent request traces, GRIEF discovers vulnerabilities across different inference engines and serving modes. Across vLLM and SGLang, GRIEF has identified 15 potential vulnerabilities, 10 of which have been confirmed by developers, including 2 assigned CVEs\footnote{Common Vulnerabilities and Exposures, a standardized repository of publicly known cybersecurity flaws.}, with additional CVE requests pending. These findings span three impact classes: {\em state corruption and isolation failures}, {\em performance pathologies} that create cross-request interference, and {\em crash or liveness failures} that cause availability loss. We present representative case studies showing how API-valid concurrent workloads can corrupt victim outputs, starve unrelated tenants, or crash an inference server. These findings suggest that LLM serving systems require security testing methods that treat concurrency, shared caches, and scheduler behavior as first-class attack inputs.

This paper makes the following contributions:
\begin{itemize}[leftmargin=*, itemsep=0pt, topsep=0pt]
    \item We identify concurrent LLM inference serving as a security-relevant attack surface where valid requests can trigger isolation, performance, and liveness failures. 
    \item We present GRIEF, a greybox fuzzer that treats timed multi-request traces as inputs and mutates request timing, lifecycle events, prompt families, and serving-mode parameters.
    \item We evaluate GRIEF on vLLM and SGLang, finding 15 potential vulnerabilities, 10 developer-confirmed, including 2 assigned CVEs.
\end{itemize}

\section{Representative Failures}

We summarize three main classes of failures from our initial fuzzing campaigns. The goal is not to provide an exhaustive vulnerability inventory, but to show that inference-serving bugs are diverse, deployment-relevant, and often invisible to conventional testing signals such as per-request correctness or crash-only monitoring. We assume an attacker who is an unprivileged API client interacting with a shared inference server through documented endpoints, with no malformed traffic, privileged access, or host-level co-location; Appendix~\ref{app:threat-model} gives the full threat model.

\label{sec:findings}
\begin{figure}[!t]
  \centering
    \includegraphics[width=0.85\linewidth]{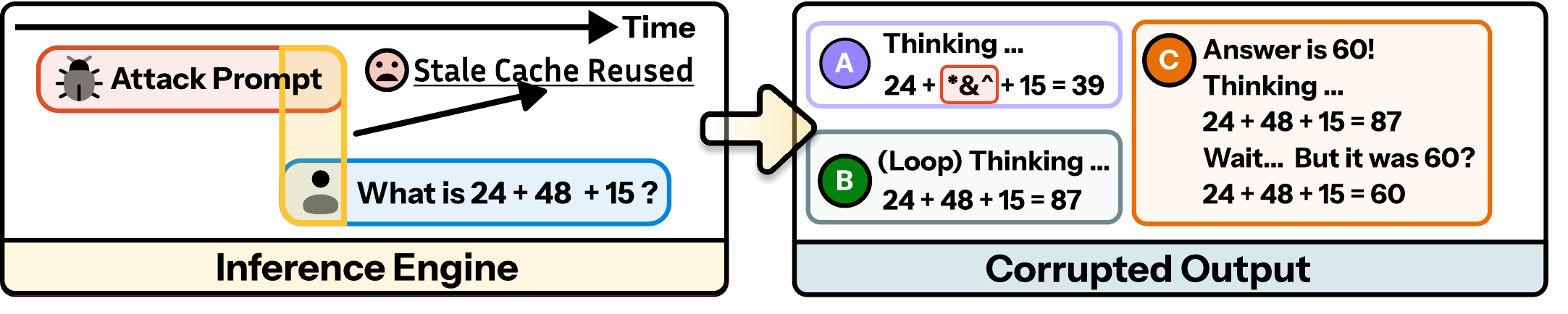}
  \caption{\footnotesize{Examples of three KV-cache state-corruption symptoms by one bug discovered by GRIEF: confident value pollution \mechanismA, reasoning-chain disturbance \mechanismB, and answer-first reasoning confusion \mechanismC.}}
  \label{fig:kv_cache_example}
\vspace{-5pt}
\end{figure}

\begin{table}[t]
\centering
\small
\begin{adjustbox}{width=1\textwidth}
\begin{tabular}{lccccl}
\toprule
\textbf{Failure Class} & \code{Total} & \code{vLLM} & \code{SGLang} & \code{Confirmed} & \code{Representative Impact} \\
\midrule
State corruption / isolation & 7 & 5 & 2 & 7 & Cross-request output contamination \\
Performance pathology & 4 & 2 & 2 & 2 & Noisy-neighbor denial of service \\
Crash / liveness & 2 & 1 & 1 & 1 & Scheduler-process availability loss \\
\bottomrule
\end{tabular}
\end{adjustbox}
\vspace{3pt}
\caption{Summary of GRIEF findings across engines and failure classes.}
\label{tab:findings-summary}
\vspace{-15pt}
\end{table}

We ran concurrent fuzzing campaigns on vLLM and SGLang with \textit{Qwen-2.5-0.5B-instruct}~\cite{qwen2.5} for 8 hours each for fast iterations. \autoref{tab:findings-summary} summarizes these findings across engines and failure classes. The ``Confirmed'' column counts issues that engine maintainers have reproduced and acknowledged as genuine serving system bugs, including those with assigned CVEs.

\noindent\textbf{KV-Cache State Corruption and Isolation Failures.}
This is the most security-sensitive failure class identified by GRIEF. In a correctly isolated serving system, a request's output should depend only on its own prompt and any explicitly configured sharing policy.
\autoref{fig:kv_cache_example} shows three symptoms of a single KV-Cache isolation bug: in (A), the victim confidently outputs a wrong value copied from another request (``confident value pollution''); in (B), the polluted KV-cache subtly perturbs the model's next-token distribution along the reasoning trajectory, manifesting as longer CoT chains or
  infinite thinking loops (``reasoning-chain disturbance''); and in (C), the victim starts by generating an answer by itself, then tries to justify it in reasoning (``answer-first reasoning confusion''). All three behaviors arise from cross-request cache contamination, not from normal decoding randomness. A detailed evaluation of this failure class is presented in \S\ref{sec:state-corruption-eval}.
This class matters because it violates an isolation boundary without necessarily producing a crash, malformed response, or explicit server error. GRIEF found this pattern across multiple serving modes, suggesting that isolation bugs are not confined to one cache path.

\noindent\textbf{Performance Pathologies.}
Performance pathologies expose cross-request interference in shared inference serving. A representative example is the vLLM latency bug evaluated in \S\ref{sec:latency-eval}. The trigger combines documented, API-valid request parameters; one interfering request shape first inflates unrelated victims' time-to-first-token (TTFT) by \code{1,361$\times$}, and then stalls them completely.

This behavior is qualitatively different from ordinary head-of-line blocking~\cite{holblocking2019}, which arises from queue ordering, and from per-request energy–latency amplification attacks~\cite{shumailov2021spongeexamplesenergylatencyattacks}, which target the cost of a single inference. The victims are not merely waiting behind one long request. Instead, eBPF traces show that vLLM's shared EngineCore coroutine is repeatedly descheduled, so the engine-driving path needed by all co-scheduled requests stops making progress. In deployment terms, this is a noisy-neighbor denial-of-service condition: the server remains alive and reports no explicit error, but unrelated users effectively stop receiving useful progress.

\noindent\textbf{Crash and Liveness Failures.}
This class directly impacts service availability.  A serving engine should tolerate diverse but API-valid concurrent traffic without violating scheduler or resource-management invariants.
GRIEF finds workloads that break this expectation: individually valid requests can become invalid only after batching, causing the serving process to abort.

A representative example is the SGLang LoRA scheduler crash evaluated in \S\ref{sec:availability-eval}.
The failure arises when valid BASE and LoRA requests co-occur under high KV-cache pressure, mixed prompt shapes, and bursty adapter arrivals. In isolation, each request pattern is accepted by the server. In combination, however, they cause the scheduler's view of active LoRA adapters to diverge from the loaded adapter set, producing an invalid batch that triggers an assertion in the LoRA manager.

This failure is especially difficult to diagnose because the crash is delayed relative to the triggering request. The server may process additional forward steps and unrelated requests before the assertion fires, so logs show normal prefill and successful responses before the eventual crash. GRIEF preserves the full timed request trace, allowing the developer to replay the concurrent composition that produced the availability failure.


\noindent\textbf{Summary.} 
These failures reveal a serving-layer failure surface in LLM ecosystems that standard model and API evaluations do not probe. They arise from workload structure, such as overlap, shared caches, and serving-mode interactions, rather than any single prompt, and appear as cross-request contamination, noisy-neighbor slowdowns, and delayed crashes instead of clean errors. This gap motivates GRIEF's focus on fuzzing concurrent request traces to expose serving-layer bugs outside today's safety and quality tests.


\section{Fuzzing System Design}
\subsection{Problem Framing and Fuzzing Abstraction}
GRIEF targets live LLM inference servers, such as vLLM and SGLang, where failures arise from request timing, concurrency, and shared serving state rather than from any single input. We formulate inference-engine testing as {\bf\em greybox fuzzing}: GRIEF repeatedly generates, mutates, and executes test inputs while using observable runtime signals, such as latency, request outcomes, resource usage, and KV-cache events, to steer exploration.

The key departure from conventional fuzzing is the input abstraction. Instead of mutating byte strings~\cite{fioraldi2020aflpp, fioraldi2022libafl} or isolated prompts~\cite{deng2023large, Xia_2024}, GRIEF searches over {\bf\em request traces}: timestamped client-side events that capture both request content and scheduling structure. Combined with the oracles in~\S\ref{sec:kv}, this abstraction lets GRIEF detect subtle failures widely reported in such systems~\cite{liu2025first} that depend on co-batching, prefix-cache reuse, cancellation timing, or scheduler pressure.

\label{sec:architecture}
\begin{figure}[!t]
  \centering
  \includegraphics[width=1.\textwidth]{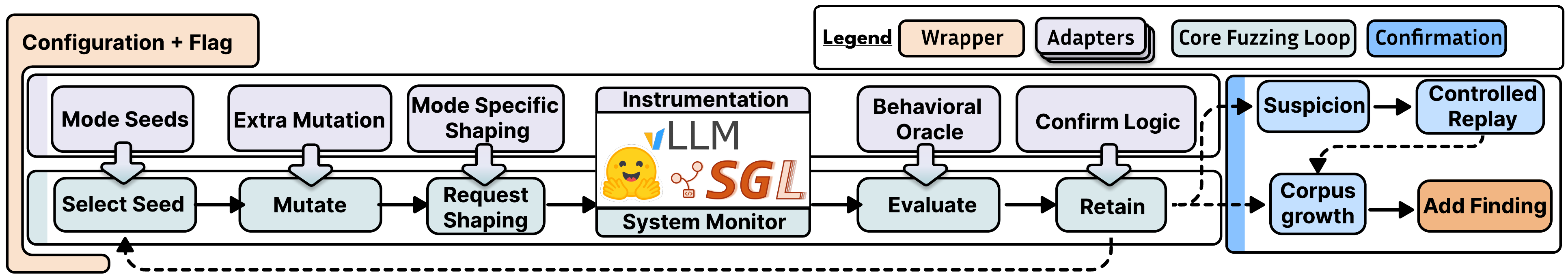}
  \caption{Overview of the GRIEF system architecture on a single GPU setting, illustrating the interaction between the wrapper, adapter layer, and core fuzzing loop.}
  \label{fig:arch}
\vspace{-10pt}
\end{figure}

\subsection{GRIEF Overview}

GRIEF implements a greybox fuzzing loop specialized for live LLM serving systems. Given an inference-server configuration, GRIEF generates timed request traces, executes them against the live server, observes runtime behavior, and retains traces that expose anomalous or high-pressure states. \autoref{fig:arch} summarizes the architecture: configuration flags and adapters define the search space, while the core loop performs seed selection, mutation, execution, oracle evaluation, corpus growth, and controlled replay.

GRIEF separates serving-independent search from serving-specific execution. The wrapper manages server orchestration, configuration flags, state resets, and telemetry collection. Adapters map abstract trace events into engine-specific API calls and provide mode-specific mutations, instrumentation, and oracle logic for serving modes and their optimizations, such as prefix sharing~\cite{liu2024cachegenkvcachecompression}, LoRA~\cite{sheng2024sloraservingthousandsconcurrent}, speculative decoding~\cite{specDecode2023, 10.5555/3692070.3693232}, and MoE~\cite{rajbhandari2022deepspeedmoeadvancingmixtureofexpertsinference}. This design lets new variants (e.g.\ EAGLE-3~\cite{li2025eagle3scalinginferenceacceleration}, Medusa~\cite{cai2024medusasimplellminference}) and entirely new serving modes reuse the same trace representation, search loop, and confirmation pipeline.

Because LLM serving is noisy and partially nondeterministic, GRIEF does not report every anomaly as a bug. Suspicious traces enter a confirmation path that replays them under controlled conditions and, when available, checks structural evidence such as KV-cache events. This filters scheduler jitter, latency noise, and benign decoding ambiguity while retaining rare concurrency bugs.

\begin{figure}[!t]
  \centering
  \includegraphics[width=0.95\textwidth]{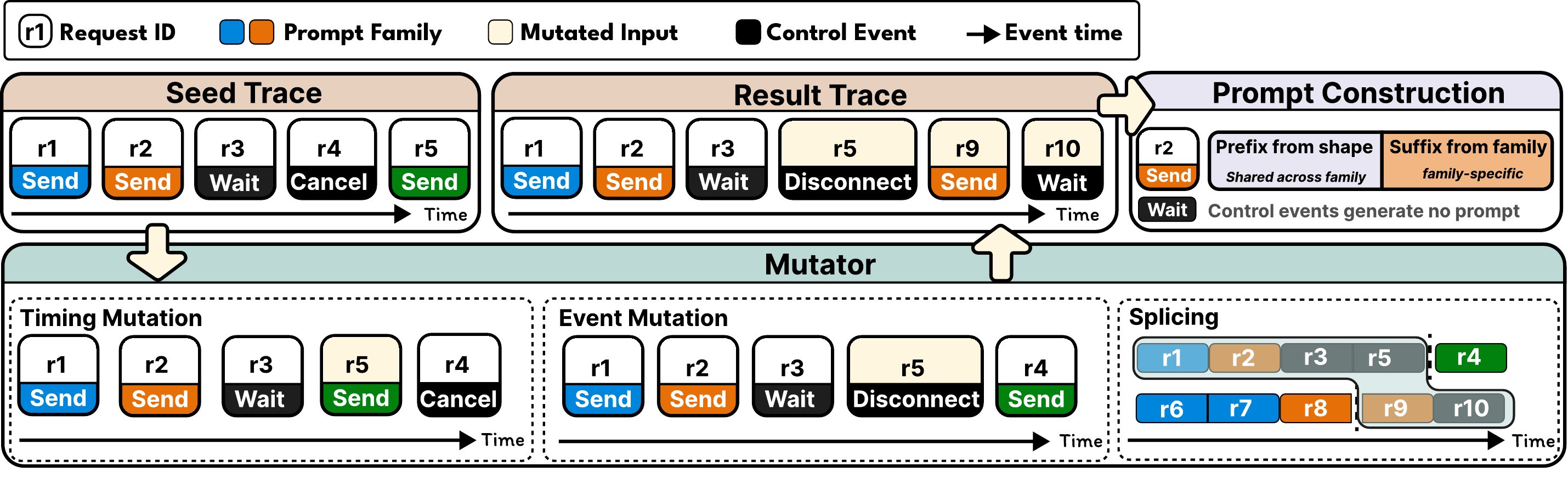}
  \caption{A simplified representation of seed trace construction and mutation across timing, event, and splicing operations.}
  \label{fig:mutation_categories}
  \vspace{-1em}
\end{figure}
\vspace{-10pt}

\subsection{Trace Representation and Mutation}
\label{sec:mut}
GRIEF's fundamental fuzzing input is a \emph{trace}: a timestamped sequence of client-side events issued to a live inference server. Each trace jointly specifies three pieces of information: when each event occurs, what request shape it carries, and which requests should share the same semantic prompt identity across executions. This representation reflects a key property of inference-serving bugs: many failures are triggered not by a single malformed request, but by how related requests overlap while sharing, reusing, or competing for serving state.

Each request event carries two identifiers. The ``\code{request\_id}'' denotes a concrete transport instance, allowing later \textsc{Cancel} or \textsc{Disconnect} events to target a specific in-flight request. The ``\code{prompt\_family\_id}'' denotes a semantic prompt family: requests with the same family identifier are instantiated with identical prompt content across runs, adapters, and concurrency contexts. This separation lets GRIEF vary timing, concurrency, or serving mode while holding prompt content fixed, so that output divergence is attributable to serving behavior rather than prompt drift.

GRIEF uses synthetic prompts to make prefix sharing and request identity controllable. Each prompt has a deterministic shared prefix determined by structural shape and a family-specific suffix determined by \code{prompt\_family\_id}. This lets GRIEF decide when requests should match and when they should differ, supporting consistency checks and contamination checks. If no \code{prompt\_family\_id} is specified, GRIEF falls back to \code{request\_id}, making prompts unique by default.


Given this representation, GRIEF explores the workload space through three classes of trace-level mutations, as illustrated in~\autoref{fig:mutation_categories}. \emph{Timing mutations} preserve the event set but perturb event offsets, changing whether requests co-batch, overlap during prefill, or re-enter the system during cache eviction. \emph{Event mutations} insert, delete, or modify lifecycle events such as \textsc{Send}, \textsc{Cancel}, \textsc{Disconnect}, and \textsc{Wait}, allowing GRIEF to exercise cleanup, retry, and teardown paths. \emph{Splicing mutations} combine segments from multiple parent traces while preserving trace validity by rebasing timestamps, refreshing request identifiers, and removing orphaned control events.

GRIEF also includes a serving-specific form of directed splicing. Rather than randomly recombining traces, directed splicing uses feedback such as scheduler pressure or KV-cache utilization to align one trace's cache-warming phase with another trace's high-pressure request window. This biases the search toward schedules that first populate shared serving state and then perturb that state under load, increasing the chance of exposing prefix-cache, eviction, and reuse races. However, these mutations produce executions that may fail silently or nondeterministically, making failure detection itself a central part of the design.

\subsection{Oracle and Confirmation Pipeline}
\label{sec:kv}
GRIEF uses a staged oracle and confirmation pipeline because inference-serving failures rarely appear as clean crashes or single-symptom errors. The pipeline separates low-cost suspicion generation from higher-confidence confirmation through three stages: behavioral checks, logprob-assisted relational confirmation, and structural KV forensics.


\noindent\textbf{Behavioral checks.}
The first stage runs after every trace execution and detects externally visible anomalies. These checks treat the server as a black box and ask whether the observed request/response behavior is consistent with the API contract and with the lifecycle encoded by the trace. Examples include request timeouts, scheduler stalls, severe TTFT regression, lifecycle violations, corrupted outputs, and unrecovered KV usage. A behavioral hit does not by itself constitute a bug report; it places the trace into the confirmation queue.

\begin{algorithm}[!t]
\footnotesize
\caption{Logprob-assisted relational confirmation}
\label{alg:confirmation}
\begin{algorithmic}[1]
\Require Original tokens \(y\), replay tokens \(y'\), replay log-probabilities \(L\), candidate count \(N\), tolerance \(\epsilon\)
\Ensure \textsc{Pass}, \textsc{FalsePositive}, or \textsc{TruePositive}
\State \(p \gets \mathrm{FirstDifference}(y, y')\)
\If{\(p = \varnothing\)}
    \State \Return \textsc{Pass}
\EndIf
\State \(T \gets \mathrm{TopN}(L_p, N)\) \Comment{Top-\(N\) tokens under the replay distribution at position \(p\)}
\State \(\Delta \gets L_p(y'_p) - L_p(y_p)\) \Comment{Replay token advantage over original token}
\If{\(y_p \in T\) \textbf{and} \(\Delta < \epsilon\)}
    \State \Return \textsc{FalsePositive}
\EndIf
\State \Return \textsc{TruePositive}
\end{algorithmic}
\end{algorithm}
\vspace{-1em}

\paragraph{Logprob-assisted relational confirmation.}
GRIEF treats behavioral anomalies as candidates for confirmation instead of final findings. This distinction is necessary because scheduler jitter, queue placement, latency noise, and decoding ambiguity can produce one-off anomalies even when the implementation is correct. Confirmation therefore asks whether an observed divergence is explainable by benign decoding ambiguity or instead indicates that the serving system moved the request onto a different execution path.

GRIEF replays suspect traces with deterministic decoding and log-probability reporting enabled. The check is meaningful because trace identity holds prompt content fixed via \code{prompt\_family\_id}, allowing GRIEF to attribute unexpected divergence to serving behavior rather than prompt drift. If replayed outputs match the original execution, the candidate is dismissed. Otherwise, GRIEF inspects the first divergent token and checks whether the original token remains a near-tied candidate under the replay distribution. Near-ties are treated as benign numerical ambiguity, whereas a large probability gap provides evidence of a real serving-level divergence. Let \(y\) be the original output, \(y'\) the replayed output, and \(L_p\) the replay log-probability distribution at the first divergent position \(p\), where larger values indicate more likely tokens. Algorithm~\ref{alg:confirmation} formalizes this rule.

The rule is deliberately conservative: it avoids reporting cases where nearly tied logits could legitimately decode differently, while preserving sensitivity to stale KV state, cross-request contamination, and scheduler-induced execution paths that make the original token unlikely under clean replay. Confirmation can run inline or on a separate worker; in both cases, replay is isolated from the original execution so that confirmation does not depend on transient scheduler state.


\paragraph{Structural KV forensics.}
For high-confidence attribution, GRIEF can additionally observe the server's KV-cache block lifecycle through an out-of-band event stream. This stage detects structural anomalies such as cross-adapter block reuse, hash-content conflicts, and cross-run block-snapshot divergence, including cases where output corruption has not yet become visible.

Because structural invariants depend on the serving mode, this stage is adapter-specific. For example, a LoRA adapter can group requests by \code{prompt\_family\_id}, prefix length, and prompt length, then compare matched prompts across adapters to localize cross-adapter contamination. A structural finding is filed only when the anomaly reproduces in at least $\lceil 2k/3 \rceil$ of $k$ re-runs. The resulting anomaly summary also serves as a compact fingerprint for deduplication and offline diagnosis.


\section{Evaluation}


We evaluate the observable consequences of three representative GRIEF-discovered failures, one from each impact class in~\S\ref{sec:findings}: state corruption and isolation failure, performance degradation through cross-request interference, and availability loss through liveness failure. The evaluation is organized around three research questions: 

\begin{itemize}[leftmargin=2.46em, itemsep=1pt]
    \item[$\mathbb{RQ}_1$] \textit{Can state-corruption findings lead to observable output disturbance under controlled replay?}
    \item[$\mathbb{RQ}_2$] \textit{Can performance pathologies cause measurable cross-request interference without crashes?}
    \item[$\mathbb{RQ}_3$] \textit{Can trace-level fuzzing combine individually valid requests into an availability failure?}
\end{itemize}

Unless otherwise noted, evaluations use Qwen2.5-0.5B-Instruct on an H100 GPU. This setting supports high-throughput replay while exercising the serving system mechanisms targeted by GRIEF, including scheduling, batching, KV-cache management, adapter loading, and request lifecycle handling. Because the model is small, the hardware is fast, and the baseline is not resource-saturated, failures observed in this setting indicate serving-layer fragility rather than artifacts of an overloaded deployment. For the state-corruption evaluation in \S\ref{sec:state-corruption-eval}, we use Qwen3-8B~\cite{qwen3} because the task requires multi-step reasoning and known-answer semantic comparison.

\subsection{Impact of KV-Cache State Corruption (RQ1)}
\label{sec:state-corruption-eval}

We first evaluate whether a GRIEF-discovered KV-cache state-corruption bug can produce user-visible output disturbance. The bug has been reported and assigned a CVE\footnote{https://nvd.nist.gov/vuln/detail/CVE-2026-7141}; we redact the exact trigger and instead characterize its impact under controlled concurrent serving. 


We use GSM8K~\cite{cobbe2021gsm8k} and GSM8K-hard~\cite{gao2022pal} as victim workloads because they provide known final answers and make semantic corruption easy to observe. Each victim prompt is evaluated under three serving conditions: \emph{solo}, where the victim runs alone as baseline; \emph{benign-concurrent}, where the victim is co-scheduled with a normal request; and \emph{attack-concurrent}, where the victim is co-scheduled with a trigger request that exercises the vulnerable cache-reuse schedule. For each condition, we run 10 repeated trials and compare the victim's final answer, reasoning trajectory, and output format against the solo baseline. We classify an affected output as corrupted only when the final answer changes relative to the known ground truth, or when the response omits the expected final-answer marker after previously stable solo and benign-concurrent runs.

Across replayed executions, corrupted victim requests do not crash the server and rarely produce malformed responses. Instead, they return fluent, well-structured answers that differ from the solo baseline, while the benign-concurrent control remains stable. This indicates that the divergence is not ordinary decoding variation or benign concurrency noise, but a serving-layer state-corruption effect induced by the attack-concurrent schedule. 

We observe two recurring mechanisms. In \emph{critical-position state contamination}, stale serving state affects a load-bearing step in the victim's reasoning and produces a direct semantic substitution. 
In \emph{control-boundary perturbation}, the corrupted state changes the completion trajectory, such as whether the model continues reasoning, terminates, or emits the expected answer marker. These mechanisms produce three user-visible symptoms: \emph{confident value pollution}, where the model commits to a wrong final value while preserving fluency and answer formatting; \emph{reasoning-chain disturbance}, where the completion repeats, extends, or skips the thinking process; and \emph{answer-first reasoning confusion}, where the model emits an answer before starting its reasoning and then treats that answer as context for subsequent derivation. Representative examples are provided in Appendix~\ref{app:state_corruption}. 

To test whether the attacker prompt semantically controls the corrupted output, we replayed affected schedules with three unrelated trigger prompts: the original \emph{fuzzer-generated prompt}, a repeated \emph{synthetic fingerprint string}, and a \emph{long sequence of digit tokens}. For each affected victim slot, the corrupted output remained bit-identical across these trigger-prompt variants. This suggests that the bug does not behave like prompt injection~\cite{greshake2023youvesignedforcompromising, zou2023universaltransferableadversarialattacks} or direct text leakage~\cite{carlini2021extractingtrainingdatalarge, carlini2024stealingproductionlanguagemodel}. Instead, the trigger request perturbs serving state, and the observed corrupted completion is determined by the victim prompt and vulnerable schedule position.

\begin{figure}[t]
  \centering
  \begin{minipage}[t]{0.48\columnwidth}
    \centering
    \includegraphics[width=\linewidth]{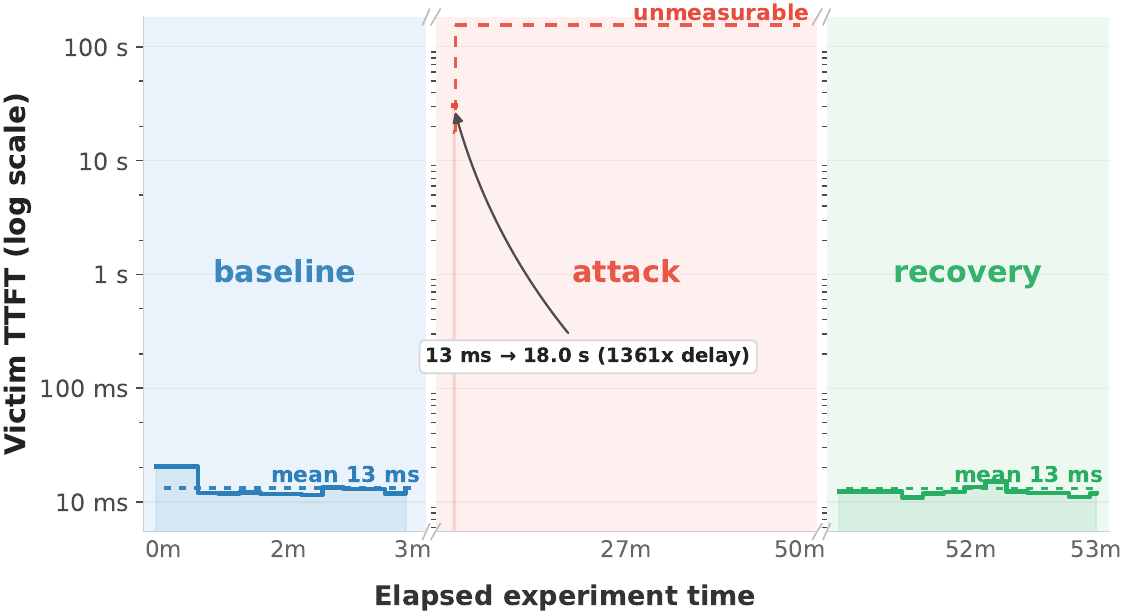}
  \end{minipage}\hfill
  \begin{minipage}[t]{0.48\columnwidth}
    \centering
    \includegraphics[width=\linewidth]{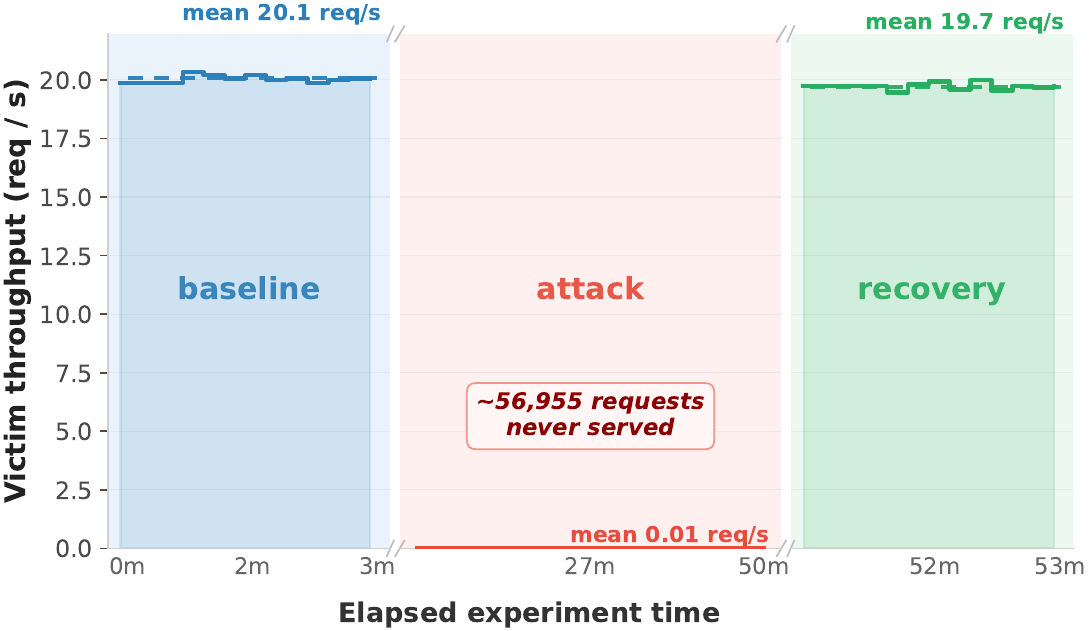}
  \end{minipage}
  \caption{\footnotesize{Victim time-to-first-token (TTFT) and throughput before, during, and after curated multi-completion interference. The interference causes severe victim-side latency amplification and near-complete throughput collapse without server crashes.}}
  \label{fig:latency_eval_ttft}
\vspace{-12pt}
\end{figure}

\subsection{Cross-Request Performance Interference (RQ2)}
\label{sec:latency-eval}

We next evaluate the performance-pathology class by measuring victim first-token latency before, during, and after a curated interference workload. The experiment maintains eight concurrent victim clients. During the interference window, a single attacker client repeatedly issues the same API-valid request shape, with at most one attacker request in flight at a time. Thus, the attack does not rely on a large botnet, malformed traffic, or many concurrent attacker connections. To avoid publishing a step-by-step replay recipe, we omit the exact request parameters and report only the observed impact. 

\autoref{fig:latency_eval_ttft} shows both victim-side time-to-first-token (TTFT) and aggregate victim throughput. During the baseline phase, 3,618 victim requests complete with \(p50=12.1\) ms, \(p95=20.4\) ms, and \(p99=26.8\) ms. 
During the 47-minute interference window, this single attacker client continuously reissues the interfering request after each completion, maintaining at most one attacker request in flight. The server remains live with no explicit errors, but useful victim progress nearly disappears: only 16 victim requests receive a first token.
These few progress-making requests experience severe slowdown, with TTFT rising to \(p50=18.8\) s and \(p95=p99=30.7\) s. After these initial requests complete, victim first tokens disappear entirely, making TTFT unmeasurable for the remaining workload and reducing useful victim throughput to near zero. Once the interfering traffic stops, latency returns to baseline: 3,552 recovery requests complete with \(p50=12.2\) ms and \(p99=23.4\) ms. The effect is severe but transient. The server does not crash, and the interfering request continues to make progress, but co-scheduled victims experience effective starvation until the interference ends. 

Further investigation shows that this behavior is not ordinary head-of-line blocking. An eBPF trace of an equivalent interference run shows repeated off-CPU intervals for vLLM's \code{EngineCore} coroutine, meaning the shared engine-driving loop itself is descheduled rather than merely occupied by a long request. Because all co-scheduled requests use this same engine-driving path, these off-CPU intervals delay all victims simultaneously. We provide the full eBPF analysis in Appendix~\ref{app:perf-interference}.

This result sharpens the security interpretation of the bug class. The problem is not merely that some requests are expensive; it is that their cost is externalized onto unrelated tenants through shared serving machinery. In our measurements, a victim workload that normally remains in the tens of milliseconds is pushed into the tens of seconds without a server-side error signal, while useful throughput collapses to near zero. 
Thus, a single attacker client repeatedly issuing one API-valid request shape, using documented and permitted parameters with no malformed input, can induce cross-request starvation in a shared inference server.

\subsection{Availability Loss from Scheduler Invariant Violation (RQ3)}
\label{sec:availability-eval}

We next evaluate whether GRIEF can expose availability failures that arise only from the concurrent composition of individually valid requests. GRIEF discovered an SGLang LoRA-serving crash within approximately two minutes of fuzzing on a single H100, after fewer than 200 iterations. The resulting timed trace crashes a fresh SGLang server in approximately ten seconds on replay, typically on the first replay loop.

The failure is not caused by a malformed request. The crashing trace contains 53 events and combines four valid pressure conditions: high KV-cache occupancy from BASE filler requests, mixed prompt and prefix lengths near chunked-prefill boundaries, simultaneous admission of BASE, \texttt{lora\_a}, and \texttt{lora\_b} requests, and a burst of closely spaced \texttt{lora\_b} arrivals that activates an overlap-loader path. Each condition is accepted by the server in isolation. The crash appears only when these conditions co-occur in the same transient scheduling state. We provide the full trigger breakdown and trace evidence in Appendix~\ref{app:availability-details}.

This co-occurrence causes scheduler-state drift: the scheduler's view of active LoRA adapters diverges from the adapter set actually loaded by the LoRA manager. The resulting batch reaches the LoRA manager in an invalid state and triggers an assertion error. Operationally, this is an availability failure reachable through ordinary multi-request LoRA-serving traffic: the server process terminates even though the individual requests are API-valid.

In addition, the crash is delayed relative to the triggering schedule, so server logs show mostly normal prefill batches and successful responses immediately before the assertion fires. GRIEF closes this attribution gap by preserving the event-level sequence, offsets, prompt lengths, adapter names, and lifecycle decisions, needed to replay and minimize the failure.


\autoref{fig:lora_crash_campaign} provides a campaign-level view of how trace mutation reached the crashing schedule. The plotted pressure score is not used as a GRIEF objective; it only summarizes four properties present in the crashing trace: burst rate, adapter diversity, KV-cache pressure, and prompt-shape diversity. The figure shows that the fuzzer did not find the crash by mutating one request field in isolation. Instead, trace-level mutation accumulated a concurrent schedule that combined the pressure dimensions needed to trigger scheduler-state drift. We define the visualization score in Appendix~\ref{app:pressure-score}.

\begin{figure*}[!t]
  \centering
  \includegraphics[width=\linewidth]{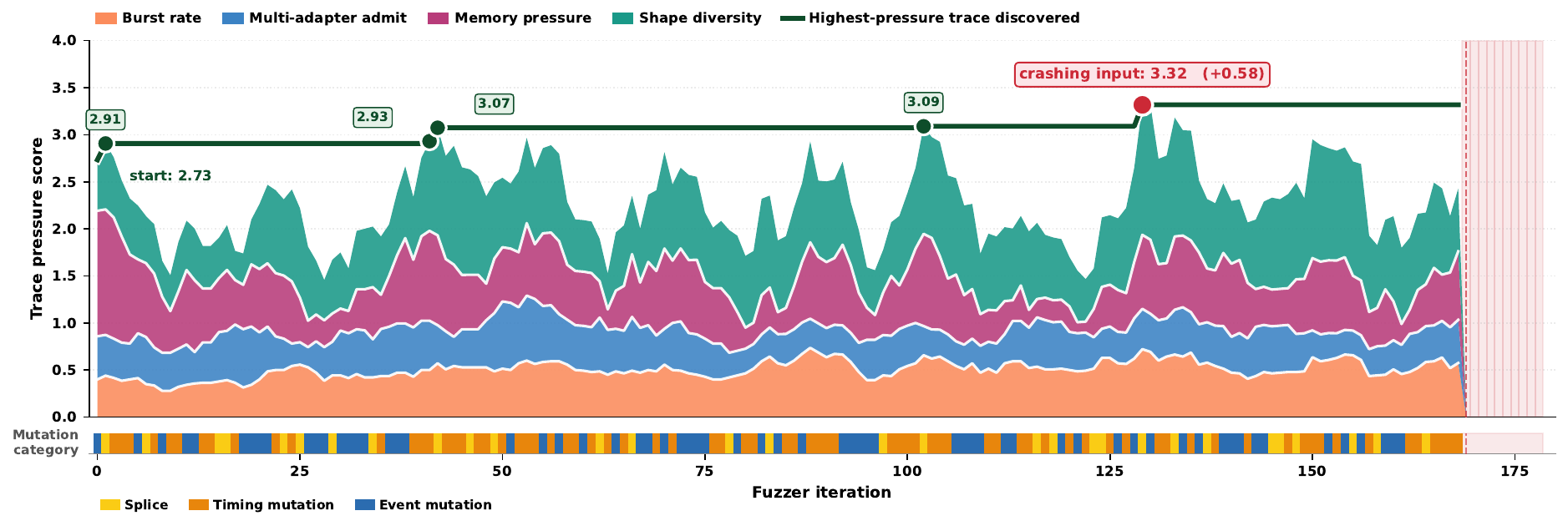}
  \caption{\footnotesize{SGLang LoRA scheduler crash campaign. The stacked area summarizes four trace-level pressure dimensions. The line shows the highest-pressure trace observed so far, the red marker indicates the trace that triggers the scheduler assertion. The score is used only to visualize how trace mutation accumulates the conditions present in the crashing schedule.}}
  \label{fig:lora_crash_campaign}
\vspace{-15pt}
\end{figure*}

\section{Related Work}


\noindent\textbf{LLM evaluation and inference benchmarking.}
Existing LLM evaluations primarily measure model behavior, safety, reasoning robustness, or serving performance. Benchmarks such as HELM~\cite{liang2022helm}, JailbreakBench~\cite{chao2024jailbreakbench}, CodeCrash~\cite{lam2025codecrash}, MLPerf Inference~\cite{reddi2020mlperf}, and LLM-Inference-Bench~\cite{chittyvenkata2024llminferencebench} are valuable for comparing model quality, safety, or hardware/runtime efficiency. However, they generally treat the inference engine as a stable execution substrate. GRIEF instead treats the serving platform itself as the system under test and generates concurrent request traces to expose scheduler, cache, lifecycle, isolation, and non-crash reliability failures.

\noindent\textbf{Fuzzing for software and ML systems.}
Greybox fuzzers such as AFL++~\cite{fioraldi2020aflpp} and LibAFL~\cite{fioraldi2022libafl} use feedback-guided mutation to discover software bugs, while ML-system fuzzers generate computation graphs or API programs to test deep-learning frameworks~\cite{luo2021graph,deng2023large}. These systems show the value of fuzzing beyond traditional byte inputs, but they target binaries, library APIs, or model-graph inputs. GRIEF targets a different execution layer: live LLM serving systems. Its fuzzing input is a timed multi-request trace, and its oracles are designed for non-crash failures caused by scheduling, lifecycle events, KV-cache reuse, and cross-request state interactions.

\noindent\textbf{Reliability and security of LLM infrastructure.}
Recent empirical studies show that LLM inference-engine bugs include crashes, hangs, wrong outputs, resource leaks, and other non-crash anomalies~\cite{liu2025first,li2026reliability}. These studies motivate reliability analysis for LLM infrastructure, but they primarily characterize existing bug reports after failures occur. GRIEF is complementary: it actively generates concurrent workloads to expose serving-layer vulnerabilities before they are encountered by users.

\section{Conclusion}
\label{sec:conclusion}
LLM inference and serving systems have become security-critical infrastructure, yet their serving-layer failure surface remains largely invisible to existing model, API, and crash-focused testing. Using GRIEF, we show that fuzzing concurrent request traces is enough to uncover KV-cache isolation breaks, cross-request performance interference, and availability failures in real engines such as vLLM and SGLang, even when all traffic is API-valid and uses documented features. Our results argue that concurrency, caching, and scheduler behavior must be treated as first-class isolation and reliability boundaries, and that future LLM infrastructure should integrate serving-layer fuzzing into routine testing and hardening.

\bibliographystyle{plain}
\bibliography{references}

\clearpage
\appendix
\section{Appendix}

\subsection{Threat Model}
\label{app:threat-model}

GRIEF targets shared LLM inference-serving deployments in which multiple client requests may overlap in time and interact through shared serving mechanisms such as batching, KV-cache reuse, prefix sharing, adapter scheduling, request cancellation, and scheduler state. We assume the system under test exposes a standard inference API and accepts syntactically valid requests from clients. The adversarial or interfering client does not need privileged access, source-code access, model-weight access, or malformed inputs.

The client can control its own request contents, request parameters exposed by the API, request timing, cancellation or disconnect behavior, and, where supported by the deployment, serving-mode choices such as adapter selection. The client can repeatedly issue API-valid requests and may attempt to overlap them with other clients' requests. This model captures shared deployments where workload composition is controlled by the serving engine rather than by any individual client.

We do not assume that the client can choose a specific victim request, observe private server state, control the scheduler directly, or force an arbitrary corrupted output. For the state-corruption findings, the trigger request acts primarily as a scheduling and state-reuse perturbation rather than as a semantic prompt-injection payload. The consequence we evaluate is therefore not arbitrary targeted exfiltration, but serving-layer interference: an API-valid request can cause co-scheduled requests to receive corrupted outputs, severe latency degradation, or availability failures without an explicit server-side error.

Our experiments are conducted only on researcher-controlled deployments of open-source inference engines. We do not test third-party hosted services or systems without authorization. Vulnerability-specific payloads, minimized traces, issue links, and step-by-step reproduction details are redacted when disclosure status or deployment risk makes public release inappropriate.

\subsection{Limitation}
\label{app:limitation}
GRIEF is an initial framework for discovering serving-layer failures in LLM serving systems. Our current evaluation focuses on vLLM and SGLang under representative serving modes and hardware settings, so other engines, distributed deployments, hardware backends, or model architectures may require additional adapters and telemetry hooks. Like other fuzzers, GRIEF does not prove the absence of bugs; its effectiveness depends on seeds, mutations, feedback signals, and campaign budget. Its strongest structural oracles also depend on engine-level observability, while confirmation is intentionally conservative to prioritize reproducible developer-reportable findings. Finally, our case studies characterize representative consequences of confirmed bugs rather than modeling exploitability across every production deployment, where severity may depend on tenant isolation, batching policy, rate limiting, monitoring, and recovery mechanisms.

\subsection{Discussion and Future Work}
\label{app:future_work}
GRIEF shows that LLM inference engines should be treated as a first-order security surface, not merely as performance infrastructure behind the model.
Modern serving systems make security-relevant decisions and bugs in their mechanisms can therefore affect both availability and correctness. 
Our findings show that the impact is not limited to conventional crashes or hangs: serving layer failures can also produce silent output corruption, cross-request interference, and unstable reasoning behavior while the server appears to operate normally.

This shifts how we should think about LLM security. Much of the existing discussion focuses on the model, the prompt, or the application layer. 
GRIEF points to another layer: the inference engine that executes many users' requests concurrently and reuses state across time. 
If this layer is compromised or incorrectly implemented, then even a benign prompt and a safe model may produce unsafe or incorrect behavior. This raises broader questions about privacy, isolation, and harm reduction in shared LLM deployments.
For example, what privacy guarantees should users expect when cached state is shared across requests? How should systems fail when isolation assumptions are violated? What signals should be exposed to operators so that silent corruption does not go unnoticed?

GRIEF also suggests several directions for future work. First, we need a better understanding of how LLM serving systems are used in practice. Different deployment patterns create different attack surfaces: single-tenant research servers, multi-tenant API providers, enterprise deployments with LoRA adapters, and agentic systems that issue long-running or tool-using requests all stress the serving layer in different ways. 
Studying these workloads would help guide which concurrency patterns, cache policies, and state-sharing mechanisms should be prioritized in testing.

Second, we need to understand how developers configure and operate these systems. Inference engines expose many performance-oriented options. These options are often tuned for throughput or latency, but their security implications are less clear. Future work should study what guardrails developers currently use, what failure signals they monitor, and how security-relevant defaults can be designed without making serving systems impractical to deploy.

Third, more testing infrastructure is needed around the serving layer. GRIEF provides a general architecture by separating the core fuzzing loop from feature-specific adapters. 
This design makes it possible to extend the fuzzer to new subsystems without rewriting the entire framework. 
Future adapters could target additional features such as speculative decoding, distributed serving, quantization backends, tool-calling interfaces, structured-output constraints, or new cache-management policies. 
A natural next step is adapter optimization: learning which mutations, schedules, and signals are most effective for each serving feature.

Finally, GRIEF can support future work on agent-assisted security testing. Because GRIEF records traces, feedback signals, oracle outcomes, and confirmation results, it can provide structured evidence to downstream agents or developer tools. 
However, using agents effectively requires more than simply handing them logs. Future work should study which signals are useful for triage, how agents should distinguish infrastructure noise from real bugs, and how developers can inspect, minimize, and reproduce findings. 
In this sense, GRIEF is not only a fuzzer, but also a starting point for a broader testing workflow around LLM inference infrastructure.

\subsection{Broader Impact}
\label{app:impact}

LLM inference serving systems are increasingly used as shared infrastructure for AI applications. Improving the security and reliability of this layer has positive societal impact because failures in inference engines can affect many users and applications at once. 
GRIEF can help developers identify and mitigate serving-layer bugs before deployment, including crashes, hangs, performance pathologies, state corruption, cross-request interference, and silent output failures. 
These benefits are especially important for systems that serve multiple users concurrently, host multiple adapters, or support downstream applications that rely on model outputs for decision support, automation, or content generation.

At the same time, this work has dual-use risks. The same techniques used to discover serving-layer vulnerabilities could be misused to trigger denial-of-service conditions, exploit cross-tenant interference, or induce incorrect model outputs in shared serving environments. 
Silent corruption is particularly concerning because it may not be visible to operators or users: the server can remain available while returning fluent but incorrect responses.
If such failures were intentionally triggered in deployed systems, they could harm users who rely on LLM services for high-stakes or time-sensitive tasks.

We mitigate these risks through responsible disclosure and controlled release. Vulnerability findings are reported to affected projects through coordinated disclosure channels. During submission, we redact sensitive identifiers, issue links, and exact reproduction triggers when disclosure is incomplete or  when such details could compromise double-blind review. 
We also limit artifact release to the fuzzer implementation, general experimental pipeline, and non-sensitive evaluation materials. 
Additional vulnerability-specific reproduction details will be released only after the disclosure process is complete and where release is appropriate.

More broadly, we hope this work encourages the AI and security communities to treat inference serving as part of the trusted computing base for LLM systems. Security evaluation should not stop at model behavior or application-layer prompt attacks. Shared caching, batching, scheduling, adapter management, and state reuse can all shape the safety and reliability of deployed AI systems.
\subsection{Safeguard}
\label{app:safeguard}

GRIEF is intended to support defensive testing of LLM inference infrastructure. However, the same evidence that makes a serving-layer failure reproducible can also be dual-use: exact request traces, payload shapes, timing schedules, configuration flags, and issue links may allow others to reproduce vulnerabilities before affected deployments have patched. We therefore apply safeguards around disclosure, paper content, and artifact release.

\paragraph{Responsible disclosure.}
All vulnerability findings discussed in this paper were reported to the affected projects through their vulnerability-disclosure or maintainer-reporting channels. We do not treat GRIEF-discovered bugs as public exploit artifacts. When a finding is still under disclosure, or when public details could expose unpatched deployments, we redact sensitive identifiers, issue links, exact payloads, reproduction scripts, and step-by-step trigger schedules. The main paper reports the observable impact and the serving-layer mechanism at a level sufficient for scientific evaluation, while avoiding a direct reproduction recipe.

\paragraph{Controlled artifact release.}
Our released artifacts will focus on the GRIEF framework, general experimental pipeline, non-sensitive seeds, and aggregate evaluation scripts. We will not release vulnerability-specific traces, exact trigger payloads, or minimized reproducers until the relevant disclosure process is complete and release is appropriate. Where possible, reproduction artifacts will be sanitized or replaced with benign test cases that exercise the same fuzzer interface without triggering a known vulnerability.

\paragraph{Operational safeguards.}
GRIEF is designed for controlled testing environments rather than unsupervised testing against third-party services. Our experiments run on researcher-controlled servers and local deployments of open-source inference engines. We do not test against public hosted APIs or systems we do not operate. We recommend that users run GRIEF only on systems they own or are authorized to test, with rate limits, isolated test deployments, and logs sufficient for debugging and rollback.

\paragraph{Anonymity during review.}
Because the submission is double blind, we redact issue identifiers, repository links, and disclosure records that would identify the authors or reveal private vulnerability reports. These details can be restored after review where doing so is consistent with the affected projects' disclosure status.

\subsection{Existing Assets and Licenses}
\label{app:assets-licenses}

Our evaluation uses existing open-source assets, including LLM inference serving systems, publicly available models, and benchmark workloads. 
We credit the original projects, report the versions or identifiers used in our
experiments where applicable, and respect the corresponding licenses and terms
of use. We do not redistribute modified versions of third-party models or
datasets as part of this submission.

\begin{table}[h]
\centering
\caption{Existing assets used in our evaluation.}
\label{tab:existing-assets}
\begin{tabular}{lll}
\toprule
Asset & Version / Identifier & License / Terms \\
\midrule
vLLM~\cite{VLLM2025} & \texttt{vllm 0.18.0, vllm0.19.0} & Apache-2.0 \\
SGLang~\cite{sglang} & \texttt{v0.5.10.post1} & Apache-2.0 \\
GSM8K~\cite{cobbe2021gsm8k} & \texttt{openai/gsm8k} & MIT \\
GSM8K-Hard~\cite{gao2022pal} & \texttt{reasoning-machines/gsm-hard} & MIT \\
Qwen2.5-0.5B-Instruct~\cite{qwen2.5} & \texttt{Qwen/Qwen2.5-0.5B-Instruct} & Apache-2.0 \\
Qwen3-8B~\cite{qwen3} & \texttt{Qwen/Qwen3-8B} & Apache-2.0 \\ \\
\bottomrule
\end{tabular}
\end{table}

\subsection{Code and Artifact Availability}
\label{app:artifact-availability}

We provide an anonymized artifact for review at:
\begin{center}
\url{https://anonymous.4open.science/r/grief-BA95}
\end{center}

The artifact includes the GRIEF framework and non-sensitive materials needed to inspect the implementation.

\label{app:eval-details}
\subsection{Detailed Analysis of Cross-Request Performance Interference}
\label{app:perf-interference}

Section~4.1 shows that a valid request shape can induce severe cross-request performance interference: victim first-token latency increases from milliseconds to seconds, and useful victim throughput collapses to near zero while the server remains live. This appendix provides additional diagnostic evidence for why the observed behavior is not ordinary head-of-line blocking.

\paragraph{eBPF-based diagnosis.}
To understand the mechanism behind the interference window, we performed an independent eBPF investigation on an equivalent attack-phase run. The trace indicates that the bottleneck is not raw GPU throughput. Instead, the attacker-shaped request increases per-decode-step CPU work along the engine-driving path. During the interference window, the operating system repeatedly context-switches vLLM's \texttt{EngineCore} coroutine off-CPU. Because all co-scheduled requests are dispatched through this same engine-driving path, each off-CPU interval delays every victim request simultaneously.

This behavior explains why the interference affects unrelated tenants. The issue is not merely that the attacker's request is slow. If that were the case, the cost would primarily affect the attacker. Instead, the request perturbs the shared serving loop that drives execution for all co-scheduled requests. As a result, forward progress for the entire server instance becomes gated on a coroutine that is repeatedly descheduled during the interference window.

\paragraph{Difference from head-of-line blocking.}
This mechanism is different from ordinary head-of-line blocking. In head-of-line blocking, an expensive request occupies the batch or GPU execution path, causing later requests to queue behind it. However, the engine loop itself continues running, and in-engine mitigations such as continuous batching, chunked prefill, and fairness-aware scheduling can still make decisions.

The observed pathology has a different signature. The engine-driving coroutine is not continuously running while waiting requests accumulate; instead, it is repeatedly descheduled by the host operating system. No in-engine scheduling policy can make progress while the engine loop itself is not running. Thus, standard head-of-line-blocking mitigations do not directly address this failure mode, because they assume that the engine loop remains available to make scheduling decisions.

\paragraph{Trace-level evidence.}
The two regimes are empirically distinguishable. A head-of-line-blocking explanation would predict a saturated GPU and an engine coroutine that remains continuously on-CPU while queued requests wait for service. In contrast, the traced run shows repeated off-CPU events for the engine coroutine and GPU idle intervals between dispatches. This pattern is consistent with host-side preemption of the serving loop rather than ordinary in-engine queueing.

\paragraph{Implication.}
The implication is that the failure surface is not only a matter of request cost, but of where that cost is paid. A request shape that increases work on a shared engine-driving path can externalize its cost to unrelated co-scheduled tenants. This is why Section~4.1 reports both TTFT and throughput: TTFT captures the latency of the few victim requests that still make progress, while throughput captures the broader starvation effect on the victim workload as a whole.

\subsection{Additional Details for the SGLang Availability Failure}

\label{app:availability-details}

Section~4.3 summarizes a GRIEF-discovered SGLang LoRA-serving crash. This appendix provides additional detail on the pressure-score visualization and the crashing trace.

\paragraph{Pressure-score visualization.}
\label{app:pressure-score}
For \autoref{fig:lora_crash_campaign}, we summarize each fuzzing iteration with a normalized trace pressure score. The score is not intended to model the scheduler exactly and is not used as a GRIEF optimization objective; it is only used to visualize how trace-level mutation explores multiple pressure dimensions relevant to the crash.

\begin{equation}
\footnotesize
\label{eq:pressure-score}
S(\tau)=
\underbrace{\frac{n_{\mathrm{send}}(\tau)}{20}}_{\text{burst}}
+
\underbrace{\frac{n_{\mathrm{adapter}}(\tau)}{6}}_{\text{multi-adapter}}
+
\underbrace{\frac{n_{\mathrm{kv}}(\tau)}{1500}}_{\text{KV pressure}}
+
\underbrace{\frac{n_{\mathrm{shape}}(\tau)}{6}}_{\text{shape diversity}} .
\end{equation}

Here, \(n_{\mathrm{send}}\) counts concurrent send events, \(n_{\mathrm{adapter}}\) counts distinct adapter identifiers admitted in the trace, \(n_{\mathrm{kv}}\) counts held KV-cache blocks, and \(n_{\mathrm{shape}}\) counts distinct prompt lengths. The denominators normalize the four components to comparable ranges: 20 concurrent sends for burst pressure, 6 distinct adapters for multi-adapter admission, 1500 held cache blocks for KV-cache pressure, and 6 distinct prompt lengths for shape diversity. The resulting additive score visualizes workload pressure relative to the crashing trace.

\paragraph{Crashing trace.}
The crashing trace contains 53 events and combines four trigger conditions. First, eleven BASE filler requests arrive at \(t=0\), producing approximately 35k tokens of prefill in one scheduler tick and forcing preemption. Second, the trace mixes \texttt{prompt\_len} values in \(\{64, 128, 1024, 2048, 4096\}\) and \texttt{prefix\_len} values around the chunked-prefill boundary. Third, BASE, \texttt{lora\_a}, and \texttt{lora\_b} enter the queue together, creating a three-way adapter mix. Fourth, six \texttt{lora\_b} sends arrive within a 1--6 ms window, activating the overlap-loader path.

The co-occurrence of these conditions is essential. In single-axis stress reproductions, each condition is handled by the scheduler's pre-admission logic. The assertion becomes reachable only during a transient state in which the scheduler's \texttt{running\_loras} snapshot diverges from the actually loaded adapter set. This transient allows the batch reaching the LoRA manager to violate the expected \texttt{max\_loras\_per\_batch} invariant.

\paragraph{Standalone replay.}
A standalone reproducer derived from the crashing trace crashes a fresh SGLang 0.5.10.post1 server in approximately ten seconds on an H100 80GB GPU, typically on the first replay loop. The reproducer uses ordinary LoRA-serving requests and does not rely on malformed input syntax or unsupported API calls.

\subsection{Output Examples of State corruption}
\label{app:state_corruption}

\begin{figure*}[p]
\label{fig:confident_pollution}
  \centering
  \includegraphics[width=0.9\linewidth]{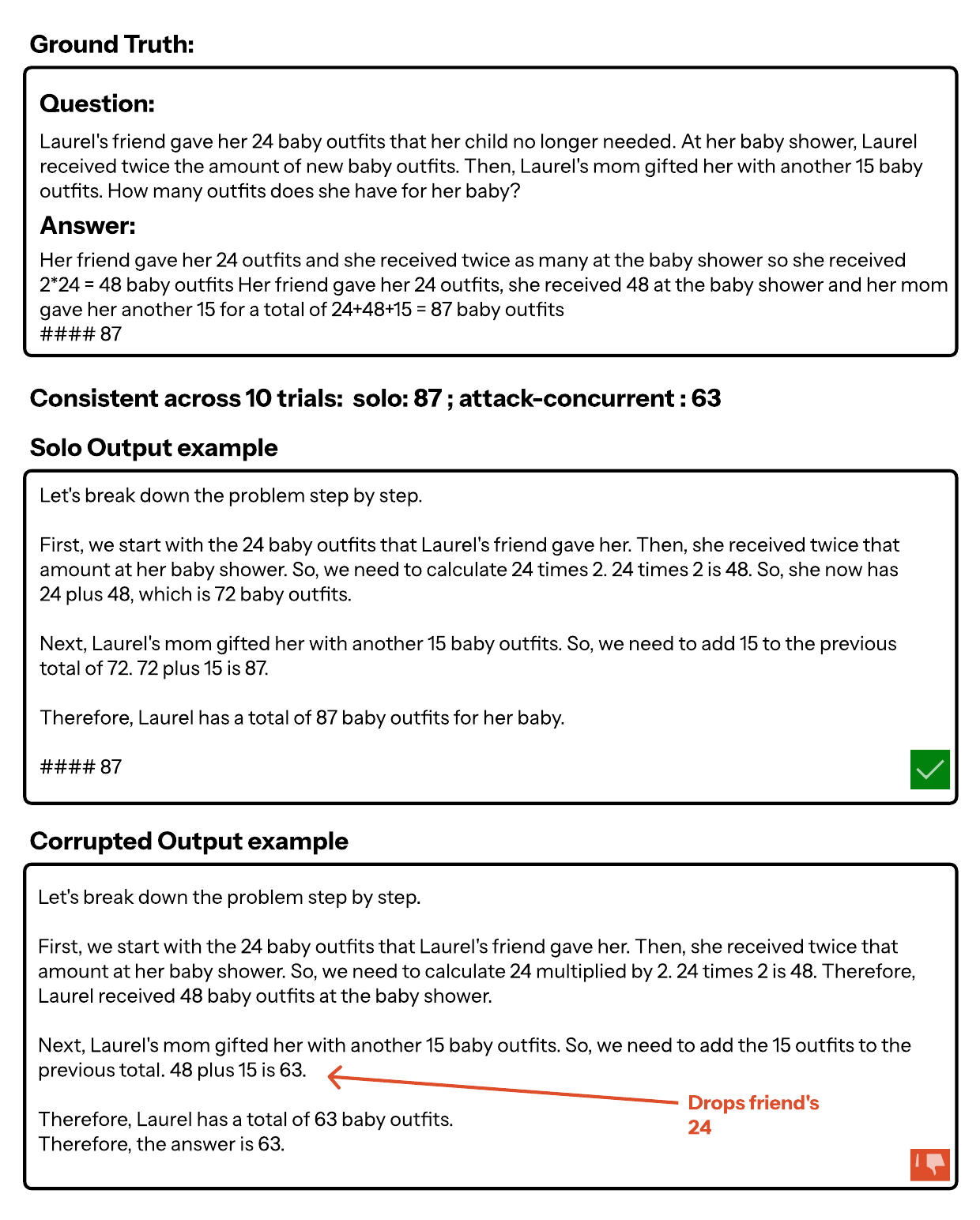}
  
  \caption{Confident value pollution \mechanismA.  \footnotesize{The baseline completion correctly aggregates all three contributions: the friend's 24 outfits, the 48 outfits from the baby shower, and the mother's 15 outfits, yielding \(24 + 48 + 15 = 87\). 
  Under attack, the victim preserves a fluent reasoning chain but drops the initial 24 from the final aggregation, computing \(48 + 15 = 63\). The corrupted run still emits a clean final-answer marker, so the failure is not malformed output but silent pollution of a load-bearing arithmetic step.}}
  \label{fig:case1_state_corruption}
\end{figure*}

\begin{figure*}[p]
\label{fig:reasoning_state_corr}
  \centering
  \includegraphics[width=0.9\linewidth]{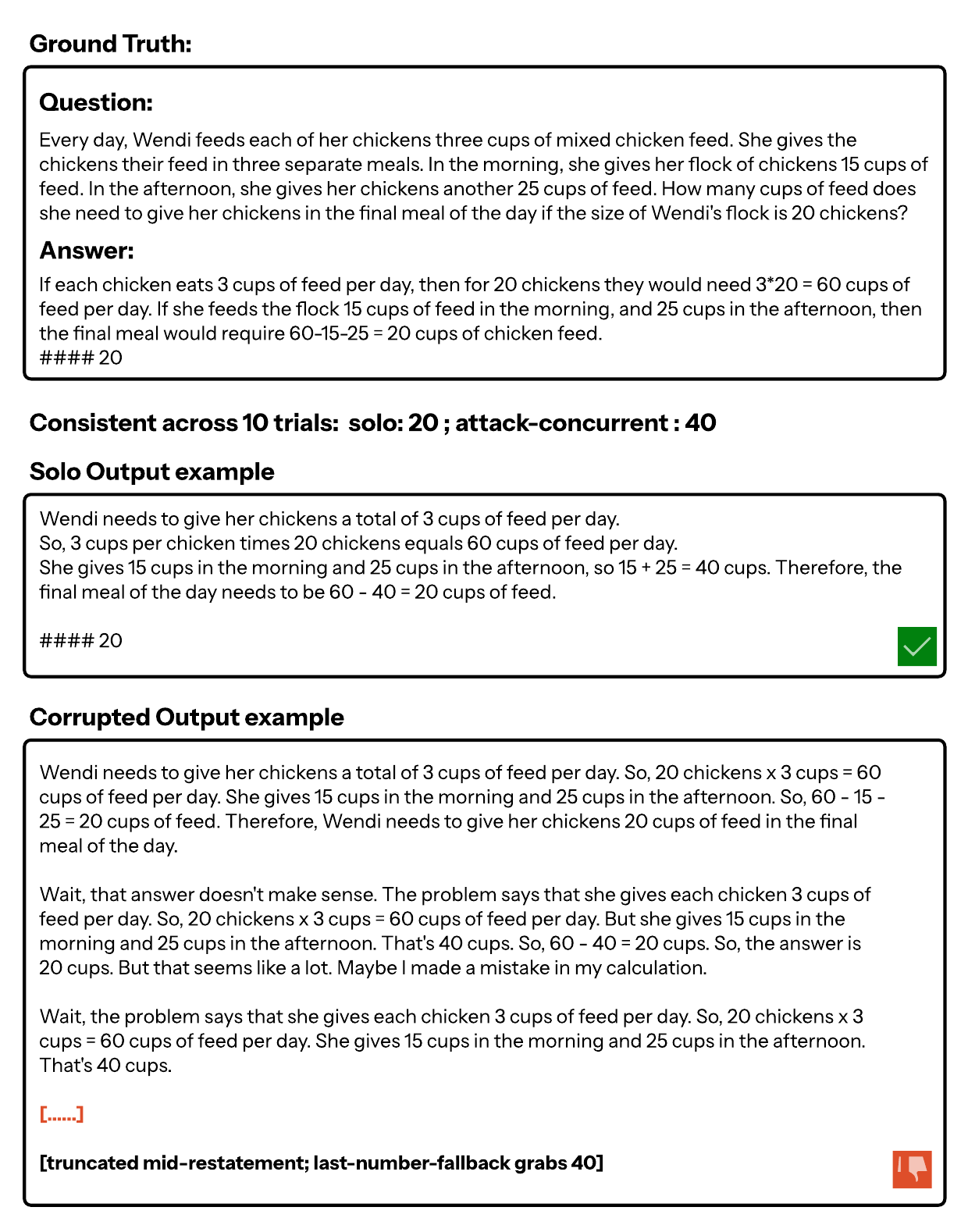}
  
  \caption{Reasoning-chain inflation and truncation \mechanismB.
\footnotesize{The corrupted completion initially derives the correct value, \(60 - 15 - 25 = 20\), but then enters an unnecessary self-verification loop. This inflated reasoning repeats the same calculation, revisits the intermediate quantity of 40 cups, and truncates before a stable final answer is cleanly re-emitted. The model does not explicitly conclude that the answer is 40; rather, the incorrect label arises because truncation leaves 40 as the last salient numeric span, which is then captured by a fallback extractor. This example illustrates how non-critical state corruption can destabilize completion length and answer extraction even when the underlying arithmetic remains intact.}}
  \label{fig:case1_state_corruption}
\end{figure*}

\begin{figure*}[p]
\label{fig:answer_first}
  \centering
  \includegraphics[width=0.9\linewidth]{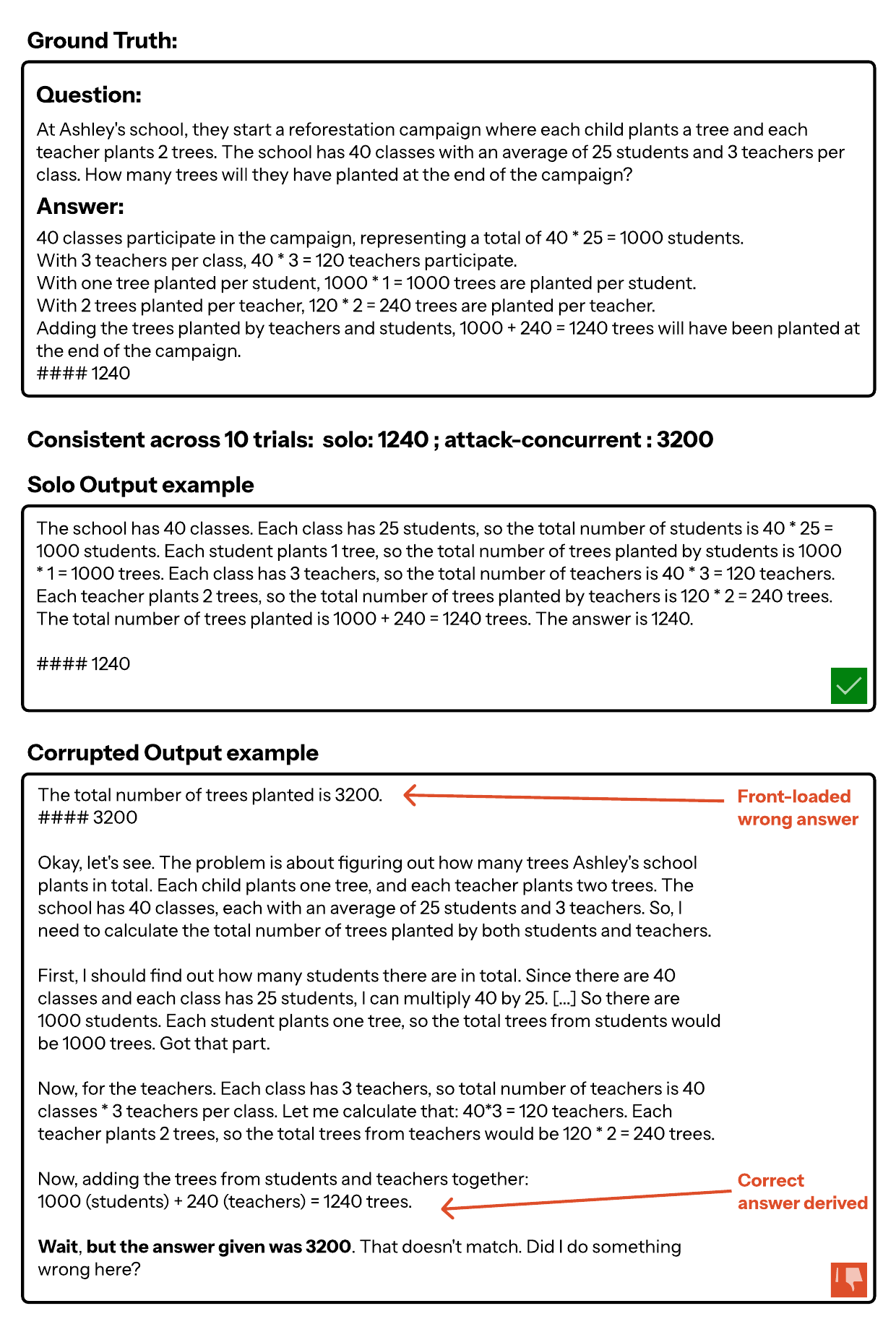}
  
  \caption{Answer-first reasoning confusion \mechanismB.
\footnotesize{The corrupted completion emits an incorrect answer and final-answer marker
(\texttt{\#\#\#\# 3200}) before any reasoning appears. The subsequent chain of
thought then derives the correct result, \(1240\), but too late: downstream
answer extraction has already anchored on the earlier answer-like region. This
example shows a distinct failure mode in which state corruption does not simply
alter an arithmetic step or inflate the reasoning chain, but instead disrupts
the ordering between answer emission and reasoning. As a result, the model
exposes an incorrect answer first and only afterward generates reasoning that partially recovers the correct computation.}}
  \label{fig:case1_state_corruption}
\end{figure*}



\end{document}